\documentclass[twocolumn,prl,showpacs]{revtex4}

\usepackage{dcolumn}
\usepackage{amsfonts}
\usepackage{amsmath}
\usepackage{amssymb}
\usepackage{bm}

\DeclareMathOperator{\tr}{tr}

\newif\ifpdf
\ifx\pdfoutput\undefined
\pdffalse 
\else
\pdfoutput=1 
\pdftrue
\fi

\ifpdf
\usepackage[pdftex]{graphicx}
\else
\usepackage{graphicx}
\fi

\begin{document}

\ifpdf
\DeclareGraphicsExtensions{.jpg,.pdf,.tif}
\else
\DeclareGraphicsExtensions{.eps,.jpg}
\fi

\newcommand{\brm}[1]{\bm{{\rm #1}}}
\newcommand{\tens}[1]{\underline{\underline{#1}}}

\title{Anomalous elasticity of nematic elastomers}

\author{Olaf Stenull}
\affiliation{
Department of Physics and Astronomy\\
University of Pennsylvania\\
Philadelphia, PA 19104\\
USA
}

\author{T. C. Lubensky}
\affiliation{
Department of Physics and Astronomy\\
University of Pennsylvania\\
Philadelphia, PA 19104\\
USA
}

\vspace{10mm}
\date{\today}

\begin{abstract}
\noindent
We study the anomalous elasticity of nematic elastomers by employing the powers of renormalized field theory. Using general arguments of symmetry and relevance, we introduce a minimal Landau-Ginzburg-Wilson elastic energy for nematic elastomers. Performing a diagrammatic low temperature expansion, we analyze the fluctuations of the displacement fields at and below the upper critical dimension 3. Our analysis reveals an anomaly of certain elastic moduli in the sense that they depend on the length scale. In $d = 3$ this dependence is logarithmic and below $d=3$ it is of power law type with anomalous scaling exponents. One of the 4 relevant shear moduli vanishes at long length scales whereas the only relevant bending modulus diverges.
\end{abstract}


\maketitle

\noindent
Nematic elastomers~\cite{FinKoc81,WarTer96,Lubensky&Co_2001} are elastic media with the macroscopic symmetry properties of nematic liquid crystals~\cite{deGennesProst93_Chandrasekhar92}.  In addition to the degrees of freedom of ordinary elastomeric materials,  liquid-crystalline elastomers possess an internal, orientational degree of freedom. For moderate crosslinking, the existence of the rubbery network has apparently little impact on the liquid crystalline phase behavior. However, because liquid-crystalline elastomers cannot flow, they have mechanical properties that differ significantly from standard liquid crystals.

Nematic elastomers have unique properties that make them candidates for device applications. Temperature change~\cite{kuepfer_finkelmann_94} or illumination~\cite{finkelmann_etal_01} can change the orientational order and cause  the elastomer to extend or contract as much as $400\%$~\cite{FinWer00}. This qualifies nematic elastomers as contestants for use in artificial muscles~\cite{hebert,ratna}.  Moreover, nematic elastomers display a soft elasticity~\cite{golubovic_lubensky_89,FinKun97_VerWar96_War99} characterized by vanishing shear stresses for a range of longitudinal strains applied perpendicular to the uniaxial direction.

The elastic energy of nematic elastomers was first studied by
Golubovi\'{c} and Lubensky (GL)~\cite{golubovic_lubensky_89}.
Normally, the elastic energy of a uniaxial elastic medium features
five independent elastic constants. However, when a nematic
elastomer is synthesized by crosslinking in the isotropic phase,
the uniaxial state arises from a spontaneous symmetry breaking at
the isotropic to nematic transition. This has the important
consequence that  the elastic constant associated with shears in
the plane containing the anisotropy axis vanishes. The work of GL
indicated that the upper critical dimension $d_c$ for nematic
elastomers is 3 or 5 in the absence or presence of random
stresses, respectively. In each case, one can expect anomalous
elasticity at or below the respective $d_c$ due to
Grinstein-Pelcovits~\cite{grinstein_pelcovits_81_82} type
renormalizations of the remaining elastic constants. Here, we
study the anomalous elasticity of nematic elastomers without
random stresses by using the powerful methods of field
theory~\cite{amit_zinn-justin} augmented by a renormalization
group (RG) analysis. Our aim is to derive the anomalous scale
dependence of the elastic moduli in and below 3 dimensions.
Parallel to our work, Xing and Radzihovsky independently studied nematic elastomers without~\cite{xing_radzihovsky0} and with~\cite{xing_radzihovsky} random stresses.
Though obtained by an entirely different formalism their final results~\cite{xing_radzihovsky0} completely agree with ours.

For our discussion, we first must devise an appropriate minimal model for nematic elastomers in form of a field theoretic Hamiltonian. Following GL, we use the Lagrangian formulation of elasticity~\cite{landau_lifshitz_elasticity_chaikin_lubensky}. In this formulation the mass points of the equilibrium undistorted medium are labeled by their position vectors $\brm{x}$ in $d$-dimensional (reference) space. When the medium is distorted, a mass point originally at $\brm{x}$ is mapped to a new point $\brm{R} (\brm{x})$ in $d$-dimensional (target) space. Since $\brm{R} (\brm{x}) = \brm{x}$ when there is no distortion, it is customary to introduce the phonon variable $\brm{u} (\brm{x}) = \brm{R} (\brm{x}) - \brm{x}$ that measures the deviation of $\brm{R} (\brm{x})$ from $\brm{x}$.  The energy of the distorted state relative to the reference state depends only on the relative amount of stretching $d \brm{R}^2 - d \brm{x}^2 = 2 u_{ij} d x_i d x_j $, where the Einstein convention is understood and where the $u_{ij} = \frac{1}{2} \left( \partial_i  u_j + \partial_j u_i +  \partial_i  u_k  \partial_j  u_k \right)$ with $i, j, k = 1, \ldots , d$ are the components of the familiar nonlinear Lagrangian strain tensor $\tens{u}$.  By construction  $\tens{u}$ is invariant under arbitrary rotations in target space. This feature makes the Lagrangian strain tensor an adequate variable for formulating elastic energies because all elastic media are rotationally invariant in target space. This invariance is easy to understand: in the absence of aligning fields different physical orientations of the same sample have the same energy.

We are interested in nematic elastomers crosslinked in the isotropic phase whose uniaxial symmetry arises via a spontaneous breaking of the isotropic symmetry. Since the spontaneous symmetry axis can point in any direction, our elastic energy has to be rotationally invariant not only in the target but also in the reference space. Both invariances are taken into account by writing the stretching energy as
\begin{eqnarray}
&&\mathcal{H} =  \int d^d x  \Big\{  \frac{\lambda}{2} \big( \tr \tens{u}  \big)^2 + \mu   \tr \tens{u}^2 + A_1 \big(  \tr \tens{u} \big)^3
\nonumber \\
&& + \, A_2 \tr \tens{u}  \tr \tens{u}^2  + A_3 \tr \tens{u}^3 + B_1 \big(  \tr \tens{u} \big)^4 + B_2 \big(  \tr \tens{u}^2 \big)^2
\nonumber \\
&& + \, B_3 \big(  \tr \tens{u} \big)^2 \tr \tens{u}^2 + B_4 \tr \tens{u}  \tr \tens{u}^3 +  B_5  \tr \tens{u}^4  \Big\}\, .
\end{eqnarray}
The first two expansion coefficients are known as the Lam\'{e} coefficients. Of course, terms of higher than fourth order are allowed by the symmetries of the system. However, these higher order terms turn out to be irrelevant in the RG sense and are hence neglected.

Suppose that the spontaneously uniaxially ordered elastomer is described in equilibrium by an equilibrium strain tensor $\tens{u}_0$. Without loss of generality we may assume that the anisotropy axis lies in the $\hat{\brm{e}}_d = (0, \ldots ,1)$ direction and that $\tens{u}_0$ is a diagonal matrix with the diagonal elements $u_{0aa} = u_{0\perp}$ and $u_{0dd} = u_{0\parallel}$. Here and in the following we use the convention that indices from the beginning of the alphabet ($a$, $b$, $c$) run from 1 to $d_\perp = d-1$. To describe deviations from the equilibrium configuration, we introduce the relative strain $\tens{w} = \tens{u} - \tens{u}_0$ and to expand $\mathcal{H}$ in terms of $\tens{w}$. By dropping terms that depend only on $\tens{u}_0$ we find
\begin{eqnarray}
\label{expansionHst1}
&&\mathcal{H} =  \int d^d x  \big\{  a_1 w_{dd} + a_2 w_{aa} + b_1 w_{dd}^2 + b_2 w_{dd} w_{aa}
\nonumber \\
&&+\,  b_3 w_{aa}^2 + b_4 w_{ab}^2 + b_5 w_{ad}^2 + c_1 w_{dd} w_{ad}^2 + c_2 w_{aa} w_{bd}^2
\nonumber \\
&&+\,  c_3 w_{ab} w_{ad} w_{bd}  + d_1 w_{ad}^2 w_{bd}^2 \big\} \, ,
\end{eqnarray}
where $\int d^d x = \int d^{d_\perp} x \int dx_d$ with $d_\perp = d -1$.  In (\ref{expansionHst1}) we have discarded terms that turn out to be irrelevant. The new coefficients $a_1$, $a_2$, $b_1$, and so on, depend on the old coefficients $\lambda$, $\mu$, and so forth, as well as on $u_{0\parallel}$ and $u_{0\perp}$.

Exploiting the rotational invariance in reference space we derived a set Ward identities relating the vertex functions implicit in (\ref{expansionHst1}). At zero-loop or mean-field level these Ward identities correspond to relations between the elastic constants in (\ref{expansionHst1}),
\begin{subequations}
\label{coeffRels}
\begin{eqnarray}
\label{coeffRelA}
a_1 - a_2 - s \, b_5 &=& 0 \, ,
\\
b_2 - 2 b_3 - s \, c_2 = 0 \, , &&b_2 + b_5 - 2 b_1 + s\,  c_1 = 0 \, ,
\\
b_5 - 2 b_4 - s \, c_3 = 0 \, , &&c_1 - c_2 - c_3 - 2 s \, d_1= 0 \, ,
\end{eqnarray}
\end{subequations}
where $s$ is an abbreviation for $u_{0\parallel} - u_{0\perp}$. Since $\tens{w}$ describes deviations from the equilibrium $\tens{u}_0$ its thermal average $\langle \tens{w} \rangle$ has to vanish. At zero temperature, where the mean-field approximation becomes exact, this means that the coefficients of the linear terms in (\ref{expansionHst1}) must be zero. Equation~(\ref{coeffRelA}) then leads to the observation that $b_5 =0$ for $u_{0\parallel} \neq u_{0\perp}$. At finite temperatures thermal fluctuations become important and loop corrections renormalize the elastic constants including $a_1$, $a_2$ and $b_5$. For $\langle \tens{w} \rangle$ to vanish the renormalized versions of $a_1$ and $a_2$ have to satisfy equations of state to which $a_1 = 0$ and $a_2 =0$ are the mean-field approximations.  In the following we will assume that we have chosen $a_1$ and $a_2$ appropriately so that their respective equations of state are satisfied. In other words: we assume that we expand about the true equilibrium state. Then, the Ward identity generalizing (\ref{coeffRelA}) guarantees that the renormalized $b_5$ vanishes for $u_{0\parallel} \neq u_{0\perp}$. The vanishing of the elastic constant $b_5$ has the striking consequence that nematic elastomers are soft with respect to certain deformations. This softness can be easily understood by looking at the terms of (\ref{expansionHst1}) that are leading for small deformations. Rewriting these terms in Fourier space, one sees that that the stretching energy cost is zero for phonon displacements $\widetilde{\brm{u}} (\brm{q})$ perpendicular to $\hat{\brm{e}}_d$ with momentum $\brm{q}$ parallel to $\hat{\brm{e}}_d$ and for $\widetilde{\brm{u}} (\brm{q})$ parallel to $\hat{\brm{e}}_d$ with $\brm{q}$ perpendicular to $\hat{\brm{e}}_d$.

For vanishing $a_1$, $a_2$ and $b_5$ one can rewrite $\mathcal{H}$ after a suitable rescaling of $x_d$ and $u_d$ as
\begin{eqnarray}
\label{Hamil}
&&\mathcal{H} = \int d^d x  \Big\{   \frac{C_1}{2} v_{dd}^2 + \frac{K}{2} \left(  \nabla_\perp^2 u_d \right)^2 + C_2 v_{dd} v_{aa}
\nonumber \\
&&+ \,  \frac{C_3}{2} v_{aa}^2 + C_4 v_{ab}^2  \Big\}  \, ,
\end{eqnarray}
with $v_{ab} = \frac{1}{2} (  \partial_a  u_b +  \partial_b  u_a - \partial_a  u_d \partial_b  u_d )$ and $v_{dd} =  \partial_d  u_d + \frac{1}{2}  \partial_a  u_d \partial_b  u_d$. Here, we exclusively retained terms that relevant in the RG sense. The new elastic constants $C_{...}$ are proportional to the $b_{...}$ in (\ref{expansionHst1}). Note that we have incorporated a relevant bending term featuring a bending modulus $K$. Bending is important in nematic elastomers since the pure stretching energy vanishes for soft deformations. Note also that $\mathcal{H}$ resembles the form of the stretching energy originally discussed by GL. However, we have to keep in mind that the strain $\tens{v}$ is different from $\tens{u}$. $\mathcal{H}$ reduces to the elastic energy by GL if one neglects terms of higher than second order in $\tens{u}$.

In principle we could use $\mathcal{H}$ as it stands in Eq.~(\ref{Hamil}) for our RG analysis. We find it convenient, however, to reduce the number of constants featured in the statistical weight $\exp (  - \mathcal{H}/T)$ at the onset of our calculation. To this end, we rescale $T \to \sqrt{K^3 /C_4} T$, $x_d \to \sqrt{C_4 /K}x_d$, $u_d \to \sqrt{K /C_4} u_d$ and $u_a \to (K /C_4) u_a$. After the rescaling, there remain 3 (dimensionless) couplings $\omega = C_1 / C_4$, $g = C_2 / C_4$, and $f = C_3 / C_4$ in our Hamiltonian. Note that we choose to keep $T$ explicit and to use it as the expansion parameter for our field theoretic perturbation expansion.

For our RG analysis, two scaling symmetries are of great importance. First, under a global rescaling of the coordinates $x_a \to \mu^{-1} x_a$ and  $x_d \to \mu^{-2} x_d$ with $\mu^{-1}$ having units of length, we find a scaling invariant theory provided that $u_a \to \mu^{1} u_a$ and $T \to \mu^{3-d} T$. Thus, the naive dimension of $u_a$ is 1 and that of $T$ is $\varepsilon = 3-d$. The field $u_d$ and the remaining parameters in $\mathcal{H}$ have vanishing naive dimensions. Above $d=3$ dimensions, the naive dimension of $T$ is negative, and $T$ is irrelevant whereas it is relevant below $d=3$. Hence, we identify $d_c =3$ as the upper critical dimension of nematic elastomers. Second, $\mathcal{H}$ is invariant under the transformation $u_a (x_c , x_d) \to u_a (x_c - \theta_c x_d , x_d) + \theta_a u_d (x_c , x_d)$ and  $u_d  (x_c , x_d) \to u_d (x_c - \theta_c x_d , x_d) + \theta_a x_a$ provided that the $\theta$'s are small. Note that this transformation mixes the longitudinal and the transversal fields (mixing invariance). It can be viewed as a rudiment of the rotational invariance of the original theory in target space. This mixing invariance is valuable because it  guarantees Ward identities that reduce the number of vertex functions to be calculated in perturbation theory.

Now we study fluctuations of the fields $u_a (\brm{x})$ and $u_d (\brm{x})$ by using perturbation theory augmented by renormalization group methods. As usual, we analyze vertex functions that require renormalization due to the presence of ultraviolet (UV) divergences in Feynman diagrams. To avoid infrared (IR) singularities in the Feynman diagrams, we supplement our Hamiltonian with a mass term $\frac{\tau}{2} \int d^d x  \,  u_{d}^2$. At the appropriate stage of the calculations we then sent $\tau$ to zero to recover the original situation.

In order to set up a diagrammatic perturbation expansion we have to determine its constituting elements. First, we have a Gaussian propagator that has the form of a $d \times d$ matrix. Its matrix elements are
\begin{subequations}
\begin{eqnarray}
G_{dd} &=&T \,  \frac{B}{B \tau +A q_d^2 + B \brm{q}_\perp^4}  \, ,
\\
G_{ad} &=& T\,  \frac{-g}{B \tau +A q_d^2 + B \brm{q}_\perp^4} \, \frac{q_a q_d}{\brm{q}_\perp^2} \, ,
\\
G_{ab} &=&  T \left[ \frac{\delta_{ab}}{\brm{q}_\perp^2}  - \frac{D \tau +C q_d^2 + D \brm{q}_\perp^4}{B \tau +A q_d^2 + B \brm{q}_\perp^4} \, \frac{q_a q_b}{\brm{q}_\perp^4} \right] \, ,
\end{eqnarray}
\end{subequations}
where we have used the shorthand notations $A = \omega (f+2) - g^2$, $B = f+2$, $A = \omega (f+1) - g^2$, and $B = f+1$. Second, our diagrammatic expansion features the 4 vertices $i (\omega - g)/(2 T)  \, q_d^{(1)}  q_b^{(2)} q_b^{(3)}$, $i (g - f)/(2 T)  \, q_a^{(1)}  q_b^{(2)} q_b^{(3)}$, $-i h /T \, q_a^{(2)}  q_b^{(1)} q_b^{(3)}$, and $- (\omega -2g + f+2)/(8  T)  \, q_a^{(1)}  q_a^{(2)} q_b^{(3)}  q_b^{(4)}$. It is understood that the sum of the momenta at each vertex has to be zero.

Next we need to determine which of the vertex functions $\Gamma^{(N,M)}$~\cite{amit_zinn-justin} with $N$ external $u_a$-legs and $M$ external $u_d$-legs are superficially UV divergent. Analyzing their topology, we find that the superficial degree of divergence $\delta$ of our diagrams is given at the upper critical dimension  by $\delta = 4 - N - 2D_\parallel - D_\perp$, where $D_\parallel$ ($D_\perp$) is the number of longitudinal (transversal) derivatives on the external legs. Thus, the vertex functions containing superficially divergent diagrams are $\Gamma^{(0,1)}_{d}$, $\Gamma^{(1,0)}_{a}$, $\Gamma^{(0,2)}_{dd}$, $\Gamma^{(1,1)}_{ad}$, $\Gamma^{(2,0)}_{ab}$, $\Gamma^{(0,3)}_{ddd}$, $\Gamma^{(1,2)}_{add}$, and  $\Gamma^{(0,4)}_{dddd}$. All these vertex functions have to be taken into account in the renormalization procedure. By virtue of the mixing invariance, however, there exist several relations between the vertex functions in form of Ward identities. Due to these Ward identities it is sufficient for our purposes to actually calculate the 2-point functions $\Gamma^{(0,2)}_{dd}$, $\Gamma^{(1,1)}_{ad}$, and $\Gamma^{(2,0)}_{ab}$. Once the equations of state are satisfied and the 2-point functions are renormalized, the Ward identities guarantee that the remaining vertex functions are cured of their UV divergences.

We calculate the two-point vertex functions to one-loop order using dimensional regularization. In dimensional regularization the UV divergences manifest themself in the form of poles in the deviation $ \varepsilon  = 3-d$ from the upper critical dimension. We eliminate the $\varepsilon$ poles from the vertex functions by employing the renormalization scheme
\begin{subequations}
\begin{eqnarray}
x_d \to \mathaccent"7017{x}_d  = Z^{-1/2} x_d \, , && u_d \to \mathaccent"7017{u}_d  = Z^{1/2} u_d \, ,
\\
u_a \to \mathaccent"7017{u}_a  = Z u_a \, , && T \to \mathaccent"7017{T} = Z^{1/2} Z^{-1}_T \mu^\varepsilon t  \, ,
\\
\omega \to \mathaccent"7017{\omega} =  Z^{-1} Z_T^{-1} Z_\omega \omega \, , && g \to  \mathaccent"7017{g} = Z^{-1}  Z_T^{-1} Z_g g \, ,
\\
f \to  \mathaccent"7017{f} = Z^{-1}  Z_T^{-1} Z_f f \, , &&
\end{eqnarray}
\end{subequations}
where the $\mathaccent"7017{}$ indicates unrenormalized quantities. The simplest way of determining the renormalization Z-factors is minimal subtraction. In this procedure the Z-factors are chosen so that they solely cancel the $\varepsilon$ poles and otherwise leave the vertex functions unchanged. Our Z-factors are of the structure $Z_{\ldots }(t, \omega , g, f ) =1+ t \, X_{\ldots }(\omega , g, f )/ \varepsilon$, where the $X_{\ldots }(\omega , g, f )$ are rational functions.

Next, we set up a Gell-Mann--Low RG equation (RGE) for the vertex functions by exploiting the fact that the unrenormalized theory has to be independent of the arbitrary length scale $\mu^{-1}$ introduced by renormalization. At this stage we prefer to switch from the parameters $\omega$, $g$ and $f$ to $\kappa = g/\omega = C_2/C_1$, $\rho = f/\omega = C_3/C_1$ and $\sigma = 1/\omega = C_4/C_1$. This step turns out to be helpful in studying the RG flow. Our RGE reads
\begin{eqnarray}
\left\{ \mathcal{D}_\mu - \left(  N + M/2 \right) \gamma \right\}   \Gamma^{(N,M)} \left(  \left\{  \brm{q}_\perp, q_d \right\} ; t, \kappa , \rho , \sigma, \mu  \right) = 0\, ,
\nonumber
\end{eqnarray}
where $\mathcal{D}_\mu = \mu \partial_\mu - \frac{\gamma}{2} q_d \partial_{q_d} + \beta_t \partial_t +  \beta_\kappa \partial_\kappa +  \beta_\rho \partial_\rho +  \beta_\sigma \partial_\sigma$, $\beta_{...} = \mu \partial_\mu |_0 ...$ and $\gamma_{...} = \mu \partial_\mu |_0 \ln Z_{...}$~\cite{footnoteWilson}. To solve the RGE we employ the method of characteristics. We introduce a flow parameter $\ell$ and look for functions $\bar{\mu} (\ell )$,  $\bar{Z} (\ell )$, $\bar{t} (\ell )$, $\bar{\kappa} (\ell )$,  $\bar{\rho} (\ell )$ and $\bar{\sigma} (\ell )$ determined by a set of characteristic equations. These equations specify how the parameters transform if we change the momentum scale $\mu $ according to $\mu \to \bar{\mu}(\ell)=\ell \mu$. In particular, we have
\begin{subequations}
\begin{eqnarray}
\ell \partial_\ell \,  \bar{t} &=&  - \bar{t} \, \varepsilon + \bar{t}^2\,  [  -3 \bar{\kappa}^2 - 12 \bar{\kappa} \bar{\sigma} + 2 \bar{\sigma}(3+19\bar{\sigma})
\nonumber \\
&+& \bar{\rho} (3+37\bar{\sigma}) ] / [\bar{\sigma} \Omega \Delta]  \, ,
\\
\ell \partial_\ell \, \bar{\kappa} &=& \bar{t} \, (\bar{\kappa} -1)  \, (\bar{\rho} + \bar{\sigma} - \bar{\kappa}^2) \, 4 \Omega / \Delta \, ,
\\
\ell \partial_\ell \, \bar{\rho} &=& \bar{t} \, [2 \bar{\kappa}^2 (1-\bar{\rho}) + 2\bar{\rho}^2 - 4 \bar{\kappa} \bar{\sigma} + \bar{\sigma}^2
\nonumber \\
&+& \bar{\rho} (4\bar{\sigma} -2)] \, 2 \Omega / \Delta \, ,
\\
\ell \partial_\ell \, \bar{\sigma} &=& \bar{t} \, \bar{\sigma}  ( -2 + 4 \bar{\kappa} - 2\bar{\kappa}^2 + \bar{\sigma}) \, 2 \Omega / \Delta \, ,
\end{eqnarray}
\end{subequations}
where $\Omega = \sqrt{2+\bar{\rho}/\bar{\sigma}}$ and $\Delta = 64 \pi \sqrt{\bar{\rho} + 2 \bar{\sigma} - \bar{\kappa}^2}$.

First, we consider the case $d < 3$. Being interested in the IR behavior of the theory, we study the limit $\ell \to 0$. We find that the quadruple of coupling constants  $(\bar{t}( \ell) , \bar{\kappa} (\ell ), \bar{\rho}( \ell), \bar{\sigma}( \ell) )$ flows,  in a manner such that $\bar{\kappa} (\ell ) - \kappa^\ast \sim \bar{\rho}( \ell) - \rho^\ast \sim  \bar{\sigma}( \ell)$ for small $\ell$, to an IR stable fixed point $(t^\ast , \kappa^\ast,  \rho^\ast , \sigma^\ast ) = (\frac{64}{59} \sqrt{6} \, \pi \, \varepsilon , 1, 1, 0)$. Note that this fixed point implies the existence of the following universal ratios: $C_2/C_1=1$, $C_3/C_1 = 1$ and $C_4/C_1 = 0$. These should be conveniently measurable in experiments. In addition to the stable fixed point, there is the unstable Gaussian fixed point $t^\ast =0$ and there are two unstable fixed lines. Solving the RGE near the stable fixed point, in conjunction with dimensional analysis to account for the naive dimensions, reveals  the scaling behavior of the vertex functions,
\begin{eqnarray}
\label{kleiner3}
&&\Gamma^{(N,M)} \left(  \left\{  \brm{q}_\perp, q_d \right\} ; t, \kappa , \rho , \sigma, \mu  \right)
\nonumber \\
&&= \, \ell^{-(d+1) + dN + (d+1)M}  Z( \ell)^{ - (N+1)/2}
\nonumber \\
&& \times \, \Gamma^{(N,M)} \Big(  \Big\{  \frac{\brm{q}_\perp}{\ell}, \frac{q_d}{\ell^{2} Z(\ell)^{1/2} } \Big\} ; t^\ast, \kappa^\ast , \rho^\ast , \sigma (\ell), \mu \ell \Big) \, ,
\end{eqnarray}
where $Z (\ell) \sim \ell^{- \frac{42}{59} \varepsilon}$ and $\sigma (\ell) \sim  \ell^{ \frac{4}{59} \varepsilon}$.

Now, the scaling behavior of the elastic constants can be easily deduced from the two point functions. Choosing $\ell = | \brm{q}_\perp|$ and switching back to the original variables featured in the elastic energy (\ref{Hamil}) we find for example that
\begin{eqnarray}
\Gamma^{(0,2)}_{dd} \left(  \brm{q}_\perp, q_d  \right) =  T^{-1} \left\{ C_1 \,  q_d^2 + K (\brm{q}_\perp ) \,  | \brm{q}_\perp|^{4} \right\}  \,
\end{eqnarray}
with a normal (constant) shear modulus $C_1$ and a momentum depending bending modulus $K (\brm{q}_\perp ) \sim | \brm{q}_\perp|^{- \eta_K}$ governed by the anomaly exponent $\eta_K =  \frac{38}{59} \varepsilon$. Note that  $K$ diverges at large length scales. The scaling behavior of the remaining elastic constants can be derived by analogous means. We find that $C_2$ and $C_3$ are normal whereas $C_4  (\brm{q}_\perp ) \sim | \brm{q}_\perp|^{\eta_4}$ is anomalous with $\eta_4 = \frac{4}{59} \varepsilon$, i.e., $C_4$ vanishes a large length scales. We mention that (\ref{kleiner3}) does not correctly describe $\Gamma^{(2,0)}_{ab}$ at $\brm{q}_\perp =0$ because we omitted a dangerous irrelevant bending modulus. This intricacy, however, does not effect our results for the relevant elastic moduli and will be discussed elsewhere.

Since $\varepsilon$ vanishes in $d=3$ the solutions to the characteristic equations are  no longer of power law type. The flow of the temperature, for example, is described at leading order by $\bar{t}( \ell) /t = [1 - \frac{7 \sqrt{6} \, t}{64  \pi} \ln (\ell ) ]^{-1}$ where $t = \bar{t}(1)$. Similarly, we have $Z (\ell) \sim [\bar{t}( \ell) /t]^{- \frac{42}{59} }$ and $\sigma (\ell) \sim [\bar{t}( \ell) /t]^{ \frac{4}{59}}$ at leading order.
Equation~(\ref{kleiner3}) now implies that
\begin{subequations}
\label{log}
\begin{eqnarray}
K (\brm{q}_\perp ) \sim \bigg[ 1 + \frac{7 \sqrt{6} \, t}{64 \pi} \ln \left( \mu/| \brm{q}_\perp| \right) \bigg]^{38/59} \, ,
\\
C_4 (\brm{q}_\perp ) \sim \bigg[ 1 + \frac{7 \sqrt{6} \, t}{64  \pi} \ln \left( \mu/| \brm{q}_\perp| \right) \bigg]^{-4/59} \, .
\end{eqnarray}
\end{subequations}
Of course, $C_1$, $C_2$, and $C_3$ are normal as they are in $d<3$ and $K$ and $C_4$ diverge and vanish, respectively, at long length scales. Note that Eqs.~(\ref{log}) imply the existence of a nonlinear length scale $\xi_t = \mu^{-1} \exp( \frac{64 \pi}{7 \sqrt{6} \, t})$.

In summary, we have derived a field theoretic model for nematic elastomers and explored, on this basis,  their fascinating anomalous elasticity in and below 3 dimensions. Our work underscores the intriguing properties of nematic elastomers and calls for experimental verification. We expect that our general approach is fruitful also for other liquid crystalline elastomers and that corresponding studies will appear in the future.

We acknowledge support by the Emmy Noether-Programm of the Deutsche Forschungsgemeinschaft  (OS) and the National Science Foundation under grant DMR 00-95631(TCL). We are grateful to H. K. Janssen, J. Toner, X. Xing, and L. Radzihovsky for fruitful discussion and to the latter two for pointing out an inconsistency in a preliminary version of this paper.


\begin{thebibliography}{}
\bibitem{footnoteWilson}
The $|_0$ indicates that bare quantities are kept fixed.
\end{thebibliography}
\end{document}